\documentclass[prb,twocolumn,showpacs]{revtex4}

\usepackage{graphicx}
\usepackage[centertags]{amsmath}

\newcommand{\cm}{\ensuremath{\,\mbox{cm}^{-1}}}
\newcommand{\K}{\ensuremath{\,\mbox{K}}}

\begin{document}

\title{Quantum paraelectric behavior of pyrochlore PMN}

\author{ S.~Kamba}\email{kamba@fzu.cz}
\affiliation{Institute of Physics, Academy of Sciences of the Czech Republic, v.v.i. Na
Slovance~2, 182 21 Prague~8, Czech Republic}
\author{D.~Nuzhnyy}
\affiliation{Institute of Physics, Academy of Sciences of the Czech Republic, v.v.i. Na
Slovance~2, 182 21 Prague~8, Czech Republic}
\author{S. Denisov}
\affiliation{Institute of Physics, Academy of Sciences of the Czech Republic, v.v.i. Na
Slovance~2, 182 21 Prague~8, Czech Republic}
\author{S.~Veljko}
\affiliation{Institute of Physics, Academy of Sciences of the Czech Republic, v.v.i. Na
Slovance~2, 182 21 Prague~8, Czech Republic}
\author{V.~Bovtun}
\affiliation{Institute of Physics, Academy of Sciences of the Czech Republic, v.v.i. Na
Slovance~2, 182 21 Prague~8, Czech Republic}
\author{M.~Savinov}
\affiliation{Institute of Physics, Academy of Sciences of the Czech Republic, v.v.i. Na
Slovance~2, 182 21 Prague~8, Czech Republic}
\author{J.~Petzelt} \affiliation{Institute of
Physics, Academy of Sciences of the Czech Republic, v.v.i. Na Slovance~2, 182 21
Prague~8, Czech Republic}

\author{M. Kalnberga and A. Sternberg}
\affiliation{Institute of Solid State Physics, University of
Latvia, Kengaraga str. 8, Riga, Latvia}

\date{\today}


\begin{abstract}
Pb$_{1.83}$Mg$_{0.29}$Nb$_{1.71}$O$_{6.39}$ (PMN) crystallizing in
a cubic pyrochlore structure exhibits, as the first dielectrics
with pyrochlore structure, typical feature of quantum
paraelectrics - its permittivity continuously increases on cooling
and levels off below $\sim$ 30\K\, without any signature of a
structural phase transition. Broad-band dielectric spectra do not
show any dielectric dispersion in the real part of permittivity up
to 8.8\,GHz. THz and infrared spectra reveal a soft polar optic
mode which is responsible for the temperature dependence of the
permittivity. The leveling-off of the permittivity at low
temperatures obeys the Barrett formula and the fitted vibrational
zero-point energy $\frac{1}{2}k_{B}T_{1}$ corresponds to the
measured soft mode frequency. The number of observed infrared
phonons exceeds that predicted from the factor-group analysis
which indicates that the structure is at least locally non-cubic.

\end{abstract}

\maketitle

\section{Introduction}
PbMg$_{1/3}$Nb$_{2/3}$O$_3$ crystallizing in the cubic perovskite
structure is one of the most studied dielectrics as the typical
representative of relaxor ferroelectrics. It exhibits a high
($\sim$10$^{4}$) and strongly diffused peak in the temperature
dependent permittivity whose position remarkably shifts from
245\K\, (at 100\,Hz) to 320\K\, (at
1\,GHz).\cite{viehland,bovtun06} Its crystal structure remains
cubic down to liquid He temperatures. The peculiar dielectric
properties are caused by a wide dielectric relaxation which
broadens and slows down on cooling.\cite{viehland,bovtun06} The
relaxation originates from the dynamics of polar clusters, which
develop below the Burns temperature T$_{d} \cong$
620\,K.\cite{burns83} The broad distribution of relaxation
frequencies has its origin in random fields and random forces as a
consequence of chemical disorder in the perovskite B sites
occupied by Mg$^{2+}$ and Nb$^{5+}$. The local ferroelectric
instability in the polar clusters was indicated by an unstable
polar optic phonon, which softens to T$_{d}$ and hardens above
T$_{d}$.\cite{bovtun06,wakimoto02b} Although the dielectric
properties of perovskite PbMg$_{1/3}$Nb$_{2/3}$O$_3$ were
intensively studied during the last fifty years, not all is
completely understood, particularly the complex and broad
dielectric dispersion.

The related compound Pb$_{1.83}$Mg$_{0.29}$Nb$_{1.71}$O$_{6.39}$
(PMN) with a pyrochlore structure, which frequently grows in the
perovskite PMN ceramics and thin films as a second phase and
significantly deteriorates its dielectric properties, has been
much less studied. The only report known to the authors is that by
Shrout and Swartz.\cite{shrout83} They investigated the dielectric
response up to 400\,kHz down to liquid He temperatures and
observed a diffuse maximum in the complex permittivity below 40\K.
The crystal structure of the pyrochlore PMN single crystal ($Fd3m$
space group) was determined in detail by Wakiya et
al.\cite{wakiya93} High-frequency dielectric properties, including
microwave, THz and infrared frequencies, have not been
investigated, although they could be useful for understanding the
reported diffused and frequency dependent maximum of the complex
permittivity. The present report aims to fill this gap in the
literature. We will show that our pyrochlore PMN ceramics does not
undergo a diffuse phase transition (reported in Ref.
\cite{shrout83}), but a quantum paraelectric behavior for which
the temperature dependent permittivity is caused only by anomalous
polar phonons. In this way, it represents the first quantum
paraelectrics with pyrochlore crystal structure.

\section{Experimental}

Pb$_{1.83}$Mg$_{0.29}$Nb$_{1.71}$O$_{6.39}$ pyrochlore ceramic samples were produced by
solid state reaction of mixed oxide powders described in details in Refs.
\cite{shrout83,mergen97}. PbO (99.5\%), Nb$_{2}$O$_{5}$ (99.5\%) and MgO (97\%) powders
were mixed and sintered at 880$\,{}^\circ \rm C$ for 8 hours. The pyrochlore cubic
structure was verified by the X-ray diffraction.

The dielectric response was investigated between 400 Hz and 1 MHz from 10\K\, to 730\K\,
using an impedance analyzer HP 4192A. The TE$_{0n1 }$ composite dielectric resonator
method\cite{krupka06} and network analyzer Agilent E8364B was used for microwave
measurements at 8.8 GHz in 100 - 350\K\, temperature interval. The cooling rate was
2\K/min.

Measurements at teraherz (THz) frequencies from 7\,cm$^{-1}$ to 33 cm$^{-1}$ (0.2 -
1.0\,THz) were performed in the transmission mode using a time-domain THz spectrometer
based on an amplified Ti - sapphire femtosecond laser system. Two ZnTe crystal plates
were used to generate (by optic rectification) and to detect (by electro-optic sampling)
the THz pulses. Both the transmitted field amplitude and phase shift were simultaneously
measured; this allows us to determine directly the complex dielectric response
$\varepsilon^{\ast}(\omega)$. An Optistat CF cryostat with thin mylar windows (Oxford
Inst.) was used for measurements down to 10\K.

Infrared (IR) reflectivity spectra were obtained using a Fourier
transform IR spectrometer Bruker IFS 113v in the frequency range
of 20 - 3000 cm$^{-1}$ (0.6 - 90 THz) at room temperature, at
lower temperatures the reduced spectral range up to 650 cm$^{-1}$
only was studied (transparency region of polyethylene windows in
the cryostat). Pyroelectric deuterated triglicine sulfate
detectors were used for the room temperature measurements, while
more sensitive liquid-He-cooled (1.5 K) Si bolometer was used for
the low-temperature measurements. Polished disk-shaped samples
with a diameter of 8 mm and thickness of $\sim$2 mm were
investigated.

\section{Results and discussions}

\begin{figure}
  \begin{center}
    \includegraphics[width=80mm]{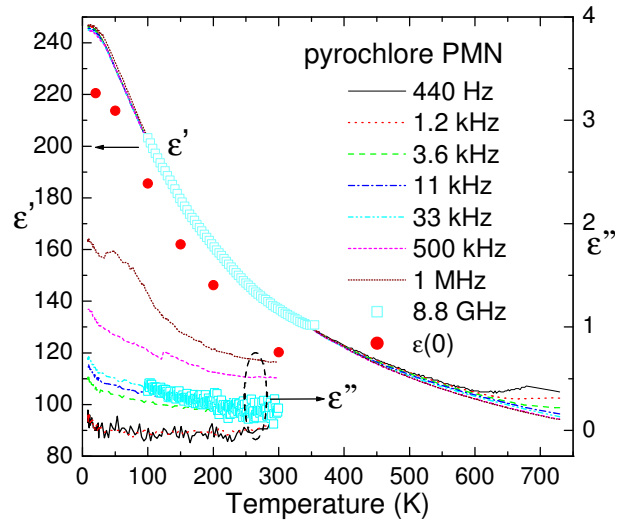}
  \end{center}
    \caption{(color online) Temperature dependence of the real $\varepsilon$' and imaginary $\varepsilon$'' part of complex permittivity
    in pyrochlore PMN ceramics at different frequencies. $\varepsilon (0)$ means the sum of phonon and
    electron contributions to the static permittivity, as
    obtained from the IR reflectivity and THz data fit. $\varepsilon$'' data are plotted only below 300\K, because
    its higher temperature values are influenced by the conductivity. Note the right scale for $\varepsilon$''.}
    \label{Fig1}
\end{figure}

\begin{figure}
  \begin{center}
    \includegraphics[width=80mm]{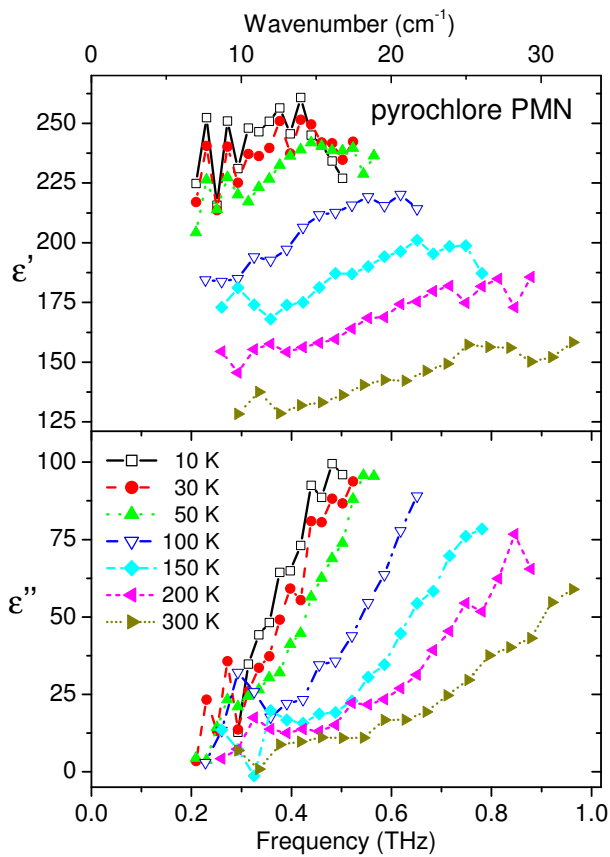}
  \end{center}
    \caption{(color online) THz dielectric spectra of the pyrochlore PMN at various temperatures. The soft mode
    frequency shifts down into the THz range on cooling, therefore the sample becomes less transparent at low temperatures
    (the noise increases) and the accessible spectral range narrows on cooling. Two frequency scales (THz and \cm) are given.}
    \label{Fig2}

\end{figure}

\begin{figure}
  \begin{center}
    \includegraphics[width=80mm]{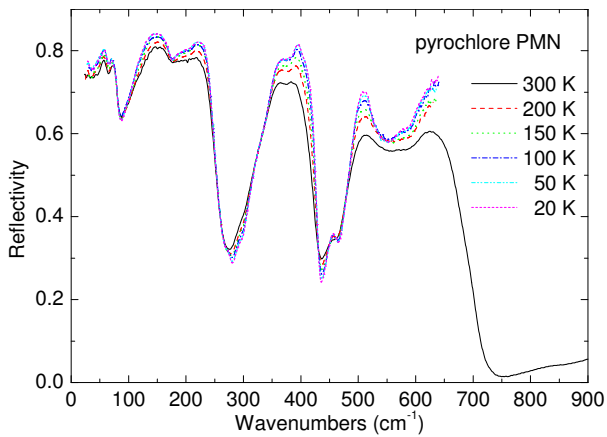}
  \end{center}
    \caption{(color online) IR reflectivity spectra of the pyrochlore PMN ceramics at various temperatures below RT.
    Low-temperature spectra were obtained only below 650\cm\, (transparency range of the polyethylene windows in the cryostat).}
    \label{Fig3}
\end{figure}

Temperature dependence of the real and imaginary parts of complex
permittivity $\varepsilon^{\ast}= \varepsilon^\prime - {\rm
i}\varepsilon^{\prime\prime}$ at various frequencies is plotted in
Fig.~\ref{Fig1}. One can see typical incipient ferroelectric
behavior, i.e. increase in $\varepsilon$' on cooling and its
noticeble saturation at low temperatures. It is important to
stress that within the accuracy of measurements no frequency
dispersion of $\varepsilon^\prime$ was observed between 400\,Hz
and 8.8\,GHz at temperatures below 600\K. The small low-frequency
dispersion above 600\K\, is caused by non-negligible conductivity
of our sample. The pronounced $\varepsilon$'(T) dependence is
therefore caused by the softening of an excitation above 10\,GHz.
To reveal it we measured the THz dielectric spectra (see
Fig.~\ref{Fig2}) and IR reflectivity spectra (Fig.~\ref{Fig3})
below room temperature (RT). One can actually see very pronounced
changes in the THz complex permittivity due to the polar phonon
softening in the investigated range (see Fig.~\ref{Fig2}).


In order to obtain all phonon parameters as a function of temperature, IR and THz spectra
were fitted simultaneously using the generalized-oscillator model with the factorized
form of the complex permittivity:\cite{gervais83}
\begin{equation}\label{eps}
\varepsilon^{*}(\omega)=\varepsilon_{\infty}\prod_{j}\frac{\omega^{2}_{LOj}-\omega^{2}+i\omega\gamma_{LOj}}{\omega^{2}_{TOj}-\omega^{2}+i\omega\gamma_{TOj}}
\end{equation}
where $\omega_{TOj}$ and $\omega_{LOj}$ denotes the transverse and
longitudinal frequency of the j-th polar phonon, respectively, and
$\gamma$$_{TOj}$ and $\gamma$$_{LOj}$ denotes their corresponding
damping constants. $\varepsilon$$^{*}$($\omega$) is related to the
normal reflectivity R($\omega$) by
\begin{equation}\label{refl}
R(\omega)=\left|\frac{\sqrt{\varepsilon^{*}(\omega)}-1}{\sqrt{\varepsilon^{*}(\omega)}+1}\right|^2
.\end{equation}

The high-frequency permittivity $\varepsilon_{\infty}$ resulting
from the electron absorption processes was obtained from the
room-temperature frequency-independent reflectivity tails above
the phonon frequencies and was assumed to be temperature
independent.

\begin{figure}
  \begin{center}
    \includegraphics[width=84mm]{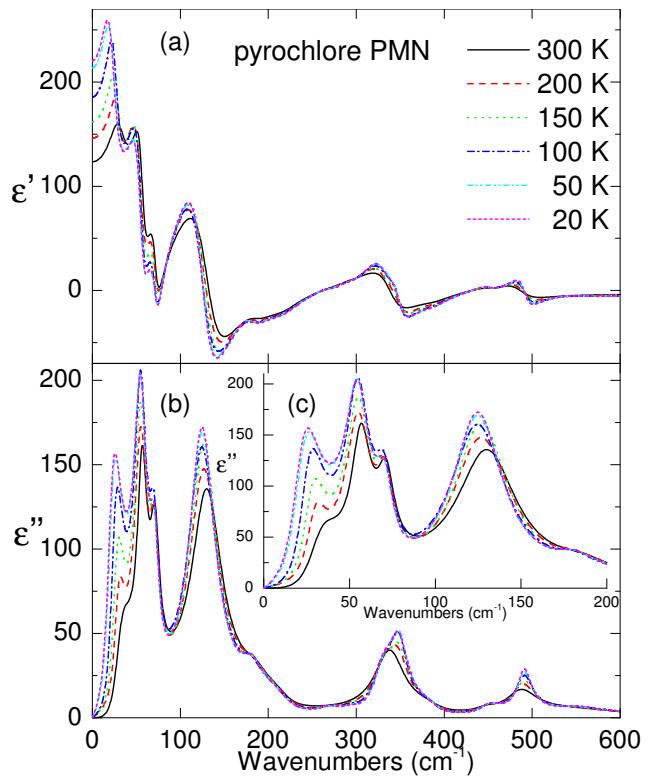}
  \end{center}
    \caption{(color online) Complex dielectric spectra from the fit to IR reflectivity and THz dielectric spectra
    at various temperatures. To see better the mode softening, $\varepsilon$''($\omega$) in inset the spectra are shown in
    the spectral range below 200\cm.}
    \label{Fig4}
\end{figure}

\begin{figure}
  \begin{center}
    \includegraphics[width=80mm]{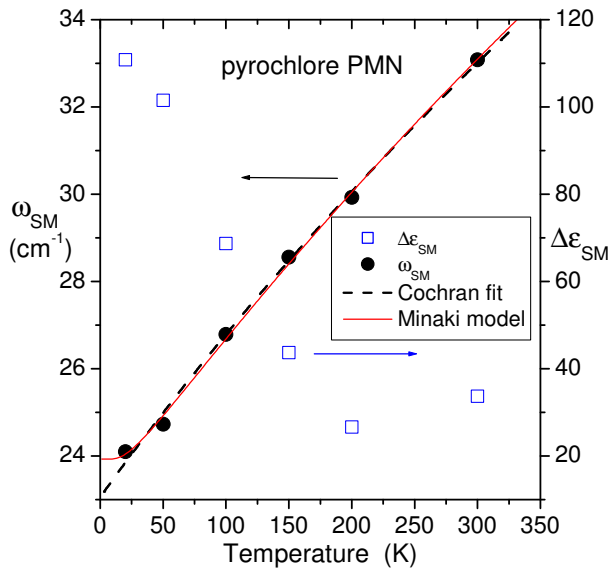}
  \end{center}
    \caption{(color online) Temperature dependence of the low-frequency soft mode $\omega_{SM}$(left scale) and its
    dielectric strength (right scale). The dashed black line and solid red line show the
    results of the Cochran and Minaki fit of the soft-mode frequency, respectively (see the text). }
    \label{Fig5}
\end{figure}

The real and imaginary parts of $\varepsilon^*$($\omega$) obtained
from the fits to IR and THz spectra are shown in Fig.~\ref{Fig4}.
Parameters of the fits performed at 300 and 20\K\, are summarized
in Table I. One can see a higher number of observed modes at
20\K\, than at RT. This is due to the reduced phonon damping at
low temperatures, which allows to resolve a higher number of
modes, which are probably overlapping at RT. One can see in the
$\varepsilon$''($\omega$) spectra of Fig.~\ref{Fig4} (we remind
that the frequencies of $\varepsilon$'' maxima roughly correspond
to the phonon frequencies) that the most remarkable frequency
shift with temperature is revealed by the lowest frequency mode
below 30\cm, but also the mode near 130\cm\, partially softens on
cooling. The temperature dependence of the soft mode frequency
$\omega_{SM}$(left scale) and its dielectric strength
$\Delta\varepsilon_{SM}$ are shown in Fig.~\ref{Fig5}. Mainly this
mode causes an increase in $\varepsilon^\prime$ on cooling (see
Fig.~\ref{Fig1}). In the case of uncoupled phonons the oscillator
strength $f_{j}=\Delta\varepsilon_{j}.\omega_{TOj}^2$ of each
phonon is roughly temperature independent so that each softening
of phonon frequency $\omega_{TOj}$ is connected with the increase
of its dielectric strength $\Delta\varepsilon_{j}$ given
by:\cite{gervais83}
\begin{equation}
 \Delta\varepsilon_{j} = \varepsilon_{\infty}\omega^{-2}_{TOj}\frac{\prod_{k}\omega^{2}_{LOk}-\omega^{2}_{TOj}}
 {\prod_{k\neq
j}\omega^{2}_{TOk}-\omega^{2}_{TOj}}.
 \label{eq:sila}
\end{equation}
In our case the soft-mode oscillator strength $f_{SM}$ is temperature dependent, it
increases twice from 3.3 .10$^{4}$ to 6.5 .10$^{4}$ cm$^{-2}$. This indicates that the
soft mode is coupled with some higher frequency mode assuming that $\sum f_{j}=const$.
However, it is difficult to reveal with which mode is the soft mode coupled, because the
oscillator strength of the most high-frequency modes is much higher than $f_{SM}$ and
relatively small changes (below the limit of our fitting accuracy) in these modes could
explain the increase of the $f_{SM}$ on cooling.

The static permittivity obtained from the fit of the IR
reflectivity is defined as
\begin{equation}\label{statics}
\varepsilon(0)=\sum_{j}\Delta\varepsilon_{j}+\varepsilon_{\infty}
\end{equation}
and is plotted in Fig.~\ref{Fig1} in red solid dots. The values of
$\varepsilon$(0) are slightly lower (by about 20) than the
experimental values obtained at and below the microwave range.
However, we believe that the disagreement is rather due to the
experimental inaccuracy than due to real dielectric dispersion
above the GHz range. So we conclude that the temperature
dependence of the permittivity below 10\,GHz in Fig.~\ref{Fig1} is
essentially due to anomalous polar phonons.

\begin{table*}[!htbp]
\caption{Parameters of the polar phonon modes in the pyrochlore
PMN obtained from the fit of IR and THz spectra at 20 and 300 K.
Frequencies $\omega_{TOj}$, $\omega_{LOj}$ and dampings
$\gamma_{TOj}$, $\gamma_{LOj}$ of modes are in
\ensuremath{\mbox{cm}^{-1}}, $\Delta \varepsilon_{j}$ is
dimensionless, $\varepsilon_{\infty}$=5.87. }
\begin{tabular}{|r | c c c c c||c c c c c |}\hline
  &\multicolumn{5}{c||}{20 K}&\multicolumn{5}{c|}{300 K}\\
  \hline
  No&$\hspace{0.2cm} \omega_{TOi} \hspace{0.2cm}$&\hspace{0.2cm}
  $\gamma_{TOj}$ \hspace{0.2cm}&\hspace{0.2cm} $\omega_{LOj}$ \hspace{0.2cm}&\hspace{0.2cm}
  $\gamma_{LOj}$ \hspace{0.2cm}&\hspace{0.2cm} $\Delta
  \varepsilon_{j}$ \hspace{0.2cm}&
  $\hspace{0.2cm} \omega_{TOj} \hspace{0.2cm}$&\hspace{0.2cm}
  $\gamma_{TOj}$ \hspace{0.2cm}&\hspace{0.2cm} $\omega_{LOj}$ \hspace{0.2cm}&\hspace{0.2cm}
  $\gamma_{LOj}$ \hspace{0.2cm}&\hspace{0.2cm} $\Delta
  \varepsilon_{j}$ \hspace{0.2cm}
 \\ \hline
1&24.1&19.4&32.5&31.5&110.8&33.2    &22.6   &37.7    &30.5   &33.7\\
 2&55.8   &16.1   &63.9   &17.7   &36.6 &56.9   &13.5   &63.2   &16.9   &23.9\\
 3&71.7   &13.4   &80.1   &14.5   &12.0 &71.9   &12.3   &79.7   &18.1   &10.8\\
 4&125.9   &36.1   &175.8   &28.4   &46.1&131.9   &42.9   &182.3   &40.6   &40.8\\
 5&179.9   &28.0   &212.0 &68.7   &1.8 &186.7   &36.2   &215.3   &56.3   &1.6\\
 6&216.7   &49.2   &257.3   &23.2   &0.8 &217.6   &41.6   &260.1   &40.2   &0.4\\
 7&271.2   &26.0   &277.3   &22.0   &0.26&&&&&\\
 8&293.5   &23.7    &296.0 &22.0   &0.2&&&&&\\
 9&329.7   &29.5   &339.8   &30.7   &0.3&&&&&\\
 10&348.9   &23.7    &388.1   &49.2   &2.0&338.8   &41.1    &383.5   &48.9   &4.3\\
 11&389.1   &36.6   &429.5   &20.6    &0.06&386.5   &40.8   &427.7   &32.4    &0.2\\
 12&449.6   &28.1    &464.1   &33.6   &0.3&448.4   &31.7   &456.3   &35.0   &0.2\\
 13&490.9   &19.9   &547.6   &59.5   &0.9&486.3   &41.9    &530.7   &116.5   &1.0\\
 14&547.8   &43.4   &551.2   &55.6   &0.001&553.1   &120 &586.1   &120.4   &0.4\\
 15&562.6   &44.8   &575.1   &70.7   &0.9&&&&&\\
 16&598.9   &70.5   &703.6   &46.8   &0.4&602.6   &104.9   &703.6   &47.6   &0.3\\
 17&862.7   &80.1   &863.0 &69.1   &0.002&862.7   &80.1   &863 &69.1   &0.002\\
  \hline
  \hline
\end{tabular}
\label{IRmodes}
\end{table*}

The temperature dependence of the lowest-mode frequency below 35\cm\, was fitted with the
Cochran law
\begin{equation}\label{Cochran}
\omega_{SM}(T)=\sqrt{A(T-T_{cr})}
\end{equation}
where A is a constant and T$_{cr}$ is the critical softening temperature. From the fit we
obtained A = (1.86$\pm$0.01) cm$^{-2}$ K$^{-1}$ and T$_{cr}$ = (-285$\pm$11)\K. So,
pyrochlore PMN tends to the ferroelectric instability at negative temperatures.

\begin{figure}
  \begin{center}
    \includegraphics[width=80mm]{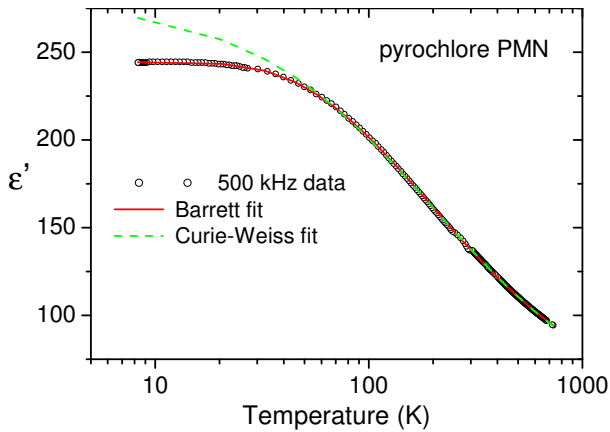}
  \end{center}
    \caption{(color online) The temperature dependence of the experimental permittivity at 500\,kHz and result of the Curie-Weiss fit
     (dashed green line)
    and the fit with Barrett formula (red solid line). Note the log temperature scale.}
    \label{Fig6}
\end{figure}

Theoretical critical temperature can be obtained also from the fit
of the temperature dependence of permittivity $\varepsilon$(T),
which (for classical paraelectrics) should follow the Curie-Weiss
law
\begin{equation}\label{CW}
 \varepsilon^\prime=\varepsilon_{cw \infty}+\frac{C}{T-T_{cw}}.
\end{equation}
The result of the Curie-Weiss fit is shown in Fig.~\ref{Fig6} in dashed line with the
following fit parameters: $\varepsilon_{cw \infty}=$43$\pm$1, Curie-Weiss constant
$C$=(47800$\pm500$)\K\, and critical temperature $T_{cw}=$(-202$\pm$8)\K. The Cochran
critical temperature T$_{cr}=$-285\K\, is somewhat lower than the Curie-Weiss critical
temperature T$_{cw}$, but if we consider that the extrapolated critical temperatures lie
far below the investigated temperature range, the agreement between both values is
reasonable.

The Curie-Weiss fit in Fig.~\ref{Fig6} deviates from the
experimental data below $\sim$ 50\K, because $\varepsilon$'(T)
levels off at low temperatures. Similar behavior was observed in
incipient ferroelectrics as SrTiO$_{3}$, KTaO$_{3}$ or CaTiO$_{3}$
where the polar soft mode is also responsible for the observed
$\varepsilon$'(T).\cite{samara01,kvyatkovskii01} The soft mode
does not soften completely, it levels off at low temperatures
(typically below 30\K), mainly due to zero-temperature vibrations
of light oxygen ions, which prevents the formation of long-range
ferroelectric order and permittivity divergence at low
temperatures. Due to this phenomenon the incipient ferroelectrics
are also called quantum paraelectrics.\cite{muller79} Note that
pyrochlore PMN is the first quantum paraelectrics of pyrochlore
crystal structure.

The leveling-off of the low-temperature permittivity in incipient
ferroelectrics was explained by Barrett\cite{barrett52} already in
the beginning of 1950's. He derived the formula
\begin{equation}\label{Barrett}
\varepsilon'(T)=\frac{M}{\frac{T_{1}}{2}\coth(\frac{T_{1}}{2T})-T_{0}}+\varepsilon_{B\infty},
\end{equation}
where M is constant, $T_{1}$ is the temperature below which the quantum fluctuations
start to play a role ($\frac{1}{2}k_{B}T_{1}$ is the zero-point vibration
energy\cite{kleemann07}) and $\varepsilon_{B\infty}$ marks the temperature independent
part of the permittivity (this term was neglected in Ref. \cite{barrett52}, because it
was very small in comparison to huge low-temperature $\varepsilon$' in SrTiO$_{3}$ and
KTaO$_{3}$). Our use of the the Barrett formula in Eq.(~\ref{Barrett}) for fitting of
$\varepsilon$'(T) yields very good agreement with the experimental data (see
Fig.~\ref{Fig6}). We found $M$=(4.25$\pm$0.01)$\times$10$^{4}$\K, $T_{1}$=(96$\pm$8)\K\,
$T_{0}$=(-167$\pm$9)\K\, and $\varepsilon_{B\infty}=$ 47$\pm$0.5. For $T\gg T_{1}$,
$\frac{1}{2}T_{1}\coth(\frac{T_{1}}{2T})$ asymptotically approaches $T$ and
Eq.(~\ref{Barrett}) becomes a Curie-Weiss law. Therefore also $\varepsilon_{B\infty}=$ 47
is close to $\varepsilon_{CW \infty}$ = 43. It is worth to note that the zero point
vibration frequency $\frac{1}{2}k_{B}T_{1}=$1\,THz=33\cm\, corresponds very well to the
soft mode frequency (see Fig.~\ref{Fig5}). Note also that the $T_{1}$ parameter as well
as the soft-mode frequency in the pyrochlore PMN qualitatively agree with analogous
parameters obtained for SrTiO$_{3}$ and SrTi$^{18}$O$_{3}$, although the values of
permittivity in these materials are two orders of magnitude
higher,\cite{barrett52,kleemann07,filipic06} which is the consequence of different $M$
parameters.

In the only published dielectric data below 400\,kHz Shrout and
Swartz\cite{shrout83} observed similar essentially dispersionless
increase in $\varepsilon$' on cooling down to 50\K\, as we did in
Fig.~\ref{Fig1}. We believe that the small dispersion below 50\K\,
and small decrease in $\varepsilon$' below $\sim$ 30\K\, observed
by Shrout and Swartz\cite{shrout83} could be due to some defects
(e.g. vacancies) in the crystal lattice of pyrochlore
Pb$_{1.83}$Mg$_{0.29}$Nb$_{1.71}$O$_{6.39}$, which is
substantially non-stoichiometric.

It is clear that the soft mode frequency cannot follow the Cochran
law (Eq.~\ref{Cochran}) in the case of low-temperature quantum
fluctuations. The correct low-temperature dependence of the soft
mode frequency derived from the Barrett formula for permittivity
can be found e.g. in the paper of Minaki et al. \cite{minaki03}

\begin{equation}\label{Minaki}
 \omega_{SM}(T)=\sqrt{A\Big[\Big(\frac{T_{1}}{2}\Big)\coth\Big(\frac{T_{1}}{2T}\Big)-{T_{0}\Big]}},
\end{equation}

where A is constant and $T_{1}$ and $T_{0}$ have the same meaning as in
Eq.(~\ref{Barrett}). Note that Eq.(~\ref{Minaki}) follows from Eq.(~\ref{Barrett}) and
Lyddane-Sachs-Teller relation under assumption that the temperature dependence of static
permittivity is caused just by softening of the soft TO mode. Result of the soft mode fit
with Eq.(~\ref{Minaki}) is shown by solid line in Fig.~\ref{Fig5} where one can clearly
see the leveling-off of the soft-mode frequency at low temperatures (unlike the Cochran
fit). The fitting parameters are the following: A = (2.03$\pm$0.05) cm$^{-2}$K$^{-1}$,
T$_{1}$ = (96$\pm$9)\K\, and T$_{0}$ = (-240$\pm$11)\K. The fit parameters could be
significantly improved if we would have more points in Fig.~\ref{Fig5} especially below
50\K. Nevertheless, one can see reasonable agreement of both Cochran and Barrett fits
above 50\K\, as well as T$_{0}$ and T$_{1}$ parameters obtained from the Barrett fits of
permittivity (Eq.~\ref{Cochran}) and the Minaki model of the soft mode frequency
(Eq.~\ref{Minaki}).

Let us compare the IR reflectivity spectra of pyrochlore PMN
(Fig.~\ref{Fig3}) with those of perovskite PMN. The latter was
first published by Burns and Dacol\cite{burns83} together with the
temperature dependence of the optical index of refraction
\textit{n}(T), which shows deviation from the linear dependence
below $\sim$ 600\K . They explained the unusual \textit{n}(T)
dependence by formation of polar nanoregions. Surprisingly, the
published IR spectrum of the perovskite sample \cite{burns83}
corresponds to our pyrochlore spectrum in Fig.~\ref{Fig3}. In
later papers of other authors (Refs.
\cite{karamyan77,reaney94,prosandeev04,hlinka06}) the mutually
similar (but different from Burns and Dacol's
spectra\cite{burns83}) IR reflectivity spectra of perovskite PMN
consisted of three distinct reflection bands typical for all cubic
perovskite oxides. We stress that the infrared spectra in Refs.
\cite{karamyan77,reaney94,prosandeev04,hlinka06} were obtained
independently on ceramics, single crystals as well as on thin
films. It appears that the IR spectrum by Burns and Dacol
\cite{burns83} belongs to pyrochlore PMN. The rest of their data
(\textit{n}(T) and \textit{P}(T)) were obviously obtained on
perovskite PMN, because the peculiarities near the Burns
temperature in the perovskite PMN were later confirmed in many
experiments.

It is of interest to compare the number of observed polar modes in
the reflectivity spectra with the prediction of factor-group
analysis: pyrochlore PMN crystallizes in $Fd3m$ space group with 8
formula units per conventional unit cell,\cite{wakiya93} i.e. 2
formula units per primitive unit cell. This means that on the
whole, 66 lattice vibrational branches are expected. Pb ions are
in 16d positions while Mg and Nb ions are in 16c positions
\cite{wakiya93}, both sites having $D_{3d}$ symmetry, while the O
cations are in positions 48f and 8b of $C_{2v}^{d}$ and $T_{d}$
symmetry, respectively. The mode symmetries and their activities
in IR and Raman spectra can be obtained using standard tables
\cite{rousseau81} with the following result for the $\Gamma$-point
of the Brillouin zone (factor-group analysis):
\begin{eqnarray}
\Gamma_{Fd3m} = 3A_{2u}(-)+ 3E_{u}(-)+8F_{1u}(x)+4F_{2u} \nonumber \\
+4F_{2g}(xy,yz,xz)+A_{1g}(x^{2},y^{2},z^{2})\nonumber\\
+E_{g}(x^{2}+y^{2}-2z^{2},\sqrt{3}x^{2}-\sqrt{3}y^{2})+2F_{1g}.
 \label{eq:pyro1}
\end{eqnarray}
It means that after subtraction of 1$F_{1u}$ acoustic mode,
7$F_{1u}$ modes should be IR active, 4$F_{2g}$, 1$A_{1g}$ and
1$E_{g}$ should be Raman active, the rest of modes being silent.
Table I shows that our fit of IR spectra required 17 modes, much
more than expected. The analysis in Eq.~\ref{eq:pyro1} assumes one
effective ion in 16c positions instead of statistically
distributed Mg and Nb ions. Since the ions strongly differ in the
mass, one could expect splitting of the modes in which these ions
take part. If we take into account different Mg and Nb vibrations,
the factor group analysis yields:
\begin{eqnarray}
\Gamma_{Fd3m} = 4A_{2u}(-)+ 4E_{u}(-)+10F_{1u}(x)+5F_{2u} \nonumber \\
+4F_{2g}(xy,yz,xz)+A_{1g}(x^{2},y^{2},z^{2})\nonumber\\
+E_{g}(x^{2}+y^{2}-2z^{2},\sqrt{3}x^{2}-\sqrt{3}y^{2})+2F_{1g}.
 \label{eq:pyro2}
\end{eqnarray}
In this case 9 polar $F_{1u}$ modes are expected, which is still
less than 17 modes observed at low temperature (see Table I). If
we assume that the Pb cations and some of the oxygen anions are
dynamically disordered among more equivalent positions with the
avarage structure remaining cubic like in isostructural
Bi$_{1.5}$ZnNb$_{1.5}$O$_{7}$,\cite{levin02}, then 14 modes could
be IR active,\cite{kamba02} close to our experimental result.

The problem with the excess IR active modes is also known from the
perovskite PMN , where only 4$F_{1u}$ polar modes are allowed in
$Fm\bar{3}m$ structure, although 7 modes were
observed.\cite{prosandeev04} The excess modes in the perovskite
PMN were explained by polar clusters, which locally break the
cubic symmetry into a rhombohedral one.\cite{hlinka06} In the case
of pyrochlore structure similar local symmetry breaking, if
present, should probably be non-polar, because there is no
indication for the existence of polar clusters. There is also no
dielectric dispersion below the polar phonon range, in contrast to
the PMN perovskite, where the huge dielectric dispersion appears
just due to polar cluster dynamics.\cite{bovtun06,viehland}
Therefore it appears that new structural studies of pyrochlore PMN
are needed to detect either a dynamical disorder of some atoms in
the lattice or a non-cubic crystal symmetry.

\section{Conclusion}

Our dielectric studies of pyrochlore PMN indicate quantum
paraelectric behavior, i.e. the permittivity increase on cooling
following the Barrett formula in the whole investigated
temperature range. The permittivity shows no dispersion up to the
microwave range and its temperature dependence can be explained by
the softening of polar optic modes. The zero-point vibrational
energy $\frac{1}{2}k_{B}T_{1}=$1\,THz obtained from the Barrett
formula corresponds very well to the soft mode frequency observed
near 30\cm. It is worth to note that the pyrochlore PMN is the
first quantum paraelectrics with pyrochlore crystal structure. Our
IR spectrum of pyrochlore PMN corresponds to IR spectrum by Burns
and Dacol\cite{burns83}, whose paper concerns the perovskite PMN.
By comparing it with the later
results,\cite{karamyan77,reaney94,prosandeev04,hlinka06} it
becomes clear that Burns and Dacol published (by mistake) the IR
spectrum of the pyrochlore PMN. Since the IR experiment revealed
more modes than expected from the factor-group analysis, we
suggest that the structure should have at least locally lower
symmetry than the cubic one.

\begin{acknowledgments}
The work has been supported by the Grant Agency of the Czech Republic (Project No.
202/06/0403) and AVOZ10100520.
\end{acknowledgments}

\end{document}